\documentclass{elsart}

\usepackage{amssymb}
\usepackage{latexsym}
\usepackage{amsmath}
\usepackage{amsfonts}

\usepackage{epsfig}

\begin{document}
\journal{Physica A}
\begin{frontmatter}

%
\title{On the size-distribution of \\ Poisson Voronoi cells}
%
\author[1,2]{J\'arai-Szab\'o Ferenc}
\author[1]{Zolt\'an N\'eda\corauthref{cor1}}\corauth[cor1]{}
\ead{zneda@phys.ubbcluj.ro},

\address[1]{Department of Theoretical and Computational Physics, Babe\c{s}-Bolyai
University, str. Kog\u{a}lniceanu 1, RO-400084 Cluj-Napoca,
Romania}

\address[2]{Interdisciplinary Computer Simulation Group, KMEI, str. Tipografiei 12, RO-400101, Cluj-Napoca, Romania}

\date{April 2007}

\maketitle

\begin{abstract}
Poisson Voronoi diagrams are useful for modeling and describing various natural
patterns and for generating random lattices. Although this particular space tessellation is
intensively studied by mathematicians, in two- and three dimensional spaces there is no exact 
result known for the size-distribution of Voronoi cells. Motivated by the 
simple form of the distribution function in the one-dimensional case, a simple and compact analytical formula is proposed for approximating the Voronoi cell's size distribution function in the practically 
important two- and three  dimensional cases as well. Denoting the dimensionality of the space 
by $d$ ($d=1,2,3$) the $f(y)=Const*y^{(3d-1)/2}exp(-(3d+1)y/2)$ compact form is suggested for the normalized 
cell-size distribution function. By using large-scale computer simulations the validity of the 
proposed distribution function is studied and critically discussed.  
\end{abstract}

\begin{keyword}
Voronoi diagrams \sep Monte Carlo methods \sep cell-size distribution

\PACS 89.75.Kd \sep 02.50.Ng \sep 02.10.-v
\end{keyword}
\end{frontmatter}

\section{Introduction}

\begin{figure}
\begin{center}
\epsfysize=80mm
\epsffile{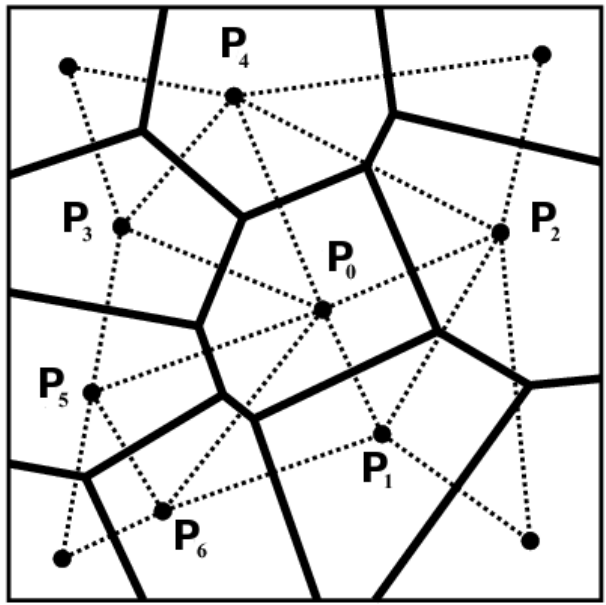}
\caption{The "perpendicular bisector method" for constructing Voronoi diagrams in 2D.}
\label{fig1}
\end{center}
\end{figure}

\begin{figure}
\begin{center}
\epsfysize=80mm
\epsffile{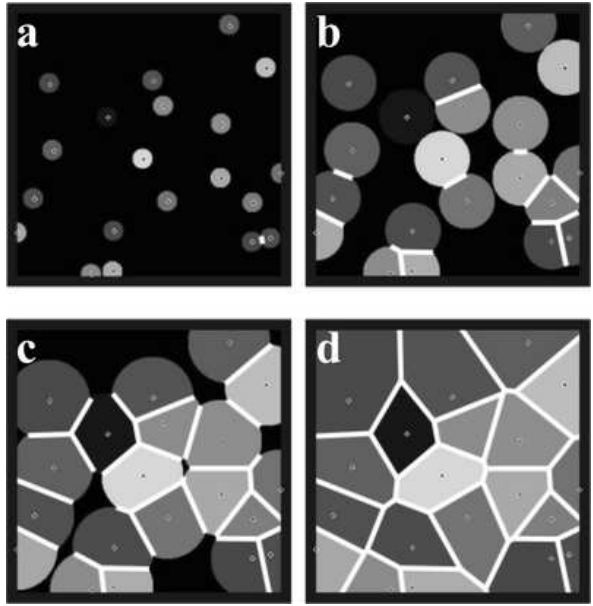}
\caption{The Avrami-Johnson-Mehl method for constructing Voronoi diagrams in 2D. Figures a.-d. presents snapshots from a small graphical simulation.}
\label{fig2}
\end{center}
\end{figure}

Voronoi diagrams \cite{book} are a particular case of space tessellation, where given a 
set of centers, the space is divided according to their "spheres of influence". Each Voronoi cell 
contains those points of the space that are closest to the same center. A Voronoi tessellation 
in two-dimension would look like the polygons sketched in Figure \ref{fig1} or Figure \ref{fig2}d. 

Given a set of centers there are two relatively easy ways to generate the corresponding 
Voronoi diagram. We sketch this methods for the two-dimensional (2D) case, and the 
generalization to any other dimension is immediate. In the {\em first method} (the perpendicular 
bisectors method \cite{book,deberg}) one starts from a given center ($P_0$) and detects the nearest 
$P_1$ center to it. A part of the perpendicular bisector on the $P_0 P_1$ line will form the first edge
of the Voronoi polygon corresponding to $P_0$. Than the second nearest center ($P_2$) 
is detected and the perpendicular bisector on $P_0P_2$ is constructed again. This algorithm 
is continued with the third ($P_3$), fourth ($P_4$), fifth ($P_5$),... nearest center, until the 
perpendicular bisectors on $P_0P_3$, $P_0P_4$, $P_0P_5$.... will close a stable polygon which does not changes 
after considering any more distant points. Repeating the above algorithm for all centers the 
Voronoi tessellation of the whole space (Figure \ref{fig1}) can be obtained.

The {\em second method} (called the Avrami-Johnson-Mehl method \cite{meijering1953}) is especially 
useful for computer simulations. In this algorithm each center is identified as a nucleation point, 
from where a virtual disc with uniform radial velocity is growing (Figure \ref{fig2}). When two discs touch each other the growth in the contact direction is stopped for both of them and the contact point becomes a point on
the corresponding Voronoi diagram. The growth in all other directions is continued until a nearby
disc is reached. In this way the same space tessellation as in the perpendicular bisector 
algorithm is achieved. In computer simulations it is handy to implement this algorithm not on continuous space 
but on large lattices since the contact points are easier to identify. 
 
In the two-dimensional case the Voronoi diagram can be obtained also from Delaunay triangulation. 
The Delaunay triangulation of a point set is a collection of edges satisfying an empty-circle criteria, which
means that for each edge we can find a circle containing the edge's endpoints but not containing any other point
from the initial set. In two-dimension the Delaunay triangulation is the dual structure of the Voronoi diagram
\cite{deberg}.

A particular case of Voronoi diagrams, where the centers are randomly and uncorrelated distributed, are called Poisson Voronoi diagrams. Poisson Voronoi Diagrams (PVD) are especially important for modeling and describing a wide variety of natural and social phenomena. PVD has been used to construct random lattices in quantum field theory \cite{Drouffe1984} or in the studies of conductivity and percolation in granular composites \cite{Jeraud}. PVD was also used in modeling growth of metal clusters on amorphous substrates \cite{DiCenzo1989}, in studying conduction and percolation in continuous media \cite{Winterfield1981}, in modeling microemulsions \cite{Talmon}, in interpreting small angle X-ray scattering for heterogeneous catalyst \cite{Brumberger1983}, in evaluating the actual galaxy distribution  \cite{Yoshioka1989}, in describing sections through various geological materials \cite{Crain1976}, in biology \cite{bio}, in animal ecology \cite{ecology}, in sociology \cite{Boots} etc... The above list is far from being complete, and suggests just a few possible applications for this particular space tessellation. For a more complete discussion of the use of Voronoi diagrams many good review works are available \cite{rivier,lecaer,weaire}.     

Despite their importance in science, our knowledge on the geometrical and statistical properties
of PVD is far from being complete \cite{book,deberg}. One of the most debated and less
clarified aspect is the $g(S)$ size distribution function of Voronoi cells $g(S)=P(S,S+dS)/dS$,
where $P(S,S+dS)$ is the probability that the size of a Voronoi cell is between $S$ and $S+dS$. Instead of $g(S)$ it is more convenient to use the more general $f(y)$ distribution function for the $y=S/ \langle S \rangle$ normalized cell sizes, which is independent of the center's density and it is universal for all PVD in a given dimension. Alternatively, one could determine the $F(y)$ cumulative distribution function defined as
$F(y)=\int_0^y f(x) dx$.

Apart of the simple one-dimensional case, presently there is no exact result or handy 
analytical approximation for the form of $f(y)$. Since $f(y)$ in two and three dimensions 
is of primary interest in many practical applications, there is a growing need for a simple and analytically usable formula. This would help characterizing and classifying several 
experimental patterns, and would give an important starting point also for modeling these structures. 
In contrast with mathematicians experimental scientist need a simple expression that could give a first 
hint about the nature of the measured cell-size distribution, which is usually determined with a poor statistics.  

There are many conjectures on the analytical form of $f(y)$ and many computer simulations were done to prove the suggested forms. Up to our knowledge in 2D the largest computer simulations were done by Tanemura 
\cite{Tanemura2003,Tanemura2004} with $10^7$ Voronoi cells and Hinde and Miles \cite{Hinde1980} 
with $2\times 10^6$ cells. In 3D the largest ensembles were studied again by Tanemura \cite{Tanemura2003,Tanemura2004} ($3 \times 10^6$ cells) and Kumar et al. \cite{kumar} ($3.6 \times 10^6$ cells).   

As a generally accepted result emerges, that a three parameter ($a$, $b$ and $c$) 
generalized gamma function fit
\begin{equation}
f(y)=c\frac{b^{a/c}}{\Gamma(a/c)}y^{a-1} exp(-by^c),
\label{gamma}
\end{equation}
describes the computer simulation data reasonable well. Some authors \cite{Weaire1986,DiCenzo1989} suggested 
however that a simpler two-parameter ($a$ and $b$) gamma function fit
\begin{equation}
f(y)=\frac{b^a}{\Gamma(a)} y^{a-1} exp(-by)
\label{fit}
\end{equation}
works also well. 

In 2D for the three parameter  fit (\ref{gamma}) Tanemura \cite{Tanemura2003,Tanemura2004} found
$a=3.315$, $b=3.04011$ and $c=1.078$, in good agreement with the 
results of Hinde and Miles \cite{Hinde1980} $a=3.3095$, $b=3.0328$ and $c=1.0787$. 
For the two parameter fit (\ref{fit}) the values $a=b=3.61$ \cite{Weaire1986} or $a=3.61$ and $b=3.57$ 
\cite{DiCenzo1989} were reported. 

In the 3D case Tanemura found \cite{Tanemura2004} $a=4.8065$, $b=4.06342$ and $c=1.16391$  
for (\ref{gamma}), while Kiang \cite{Kiang1966} suggested a fit of the form (\ref{fit}) with $a=b=6$. We have to mention however that the simulations of Tanemura \cite{Tanemura2004,Tanemura2003} did not supported 
Kiang's results \cite{Kiang1966} at all.

For the sake of completeness it also has to be mentioned that there is an exact analytical result for the 
second moment ($\langle y^2 \rangle$) of the PVD's both in the two and three-dimensional cases \cite{gilbert}.
According to this $\langle y^2_{2D} \rangle=1.280$ and  $\langle y^2_{3D} \rangle=1.180$ \cite{rivier}, offering an excellent possibility for testing the computer simulation results and the correctness of the proposed fit. 
It was the enormous discrepancy between Gilbert's and Kiang's results for this second moment that condemned Kiang's simulation results. 

The aim of the present work is not to give a better and more complicated fit for $f(y)$. We would rather intend 
to prove that a simple two parameter fit of the form (\ref{fit}) used by Kiang can be still a fair 
approximation for all practical applications. Experimental scientist instead of focusing on a more accurate but 
difficult fit for the presumed PVD type patterns can use with confidence a simple approximation 
of the form (\ref{fit}). In the present work large-scale computer simulations are also considered for 
the problem, generating more Voronoi cells than in all previous studies we are aware of. The statistics of 
$3 \times 10^7$ and $1.8 \times 10^7$ cells are studied in 2D and 3D, respectively. Using this improved 
statistics Kiang's conjecture will be followed and a first approximation for $f(y)$ in the form 
(\ref{fit}) will be given with simple and handy values of $a=b$. Solving the problem exactly in 1D will give us further motivation for this simpler form of $f(y)$.

\section{The one-dimensional case}

Let us first study theoretically the simple problem in 1D and prove the validity of 
(\ref{fit}) with $a=b$. One has to mention however that several other methods are known to
obtain the exact form of the $f_{1D}(y)$ distribution function in this simple case \cite{meijering1953,rivier}. 

\begin{figure}
\begin{center}
\epsfxsize=100mm
\epsffile{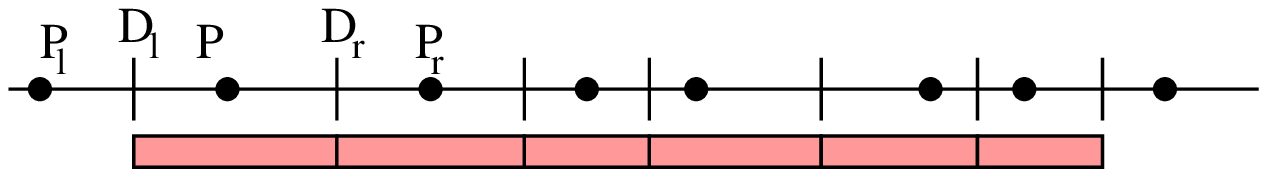}
\caption{Construction of Voronoi cells in 1D.}
\label{fig3}
\end{center}
\end{figure}

A line with length $L$ is considered on which randomly and 
independently $N$ centers are distributed. The density of centers is given thus as
$n=\frac{N}{L}=\frac{1}{\langle d \rangle}$,
where $\langle d \rangle$ stands for the average distance between centers. We will
study the limit $L\rightarrow \infty$, $N\rightarrow \infty$, but $n$ finite.
It is immediate to construct the Voronoi diagrams for these centers (Figure \ref{fig3}).
If a center $P$ is considered, first it's neighbor in the left ($P_l$) and right 
($P_r$) direction will be detected. Than the $PP_l$ and $PP_r$ lines are divided 
in two equal parts, by the
$D_l$ and $D_r$ points, respectively. The segment $D_lD_r$ is than the Voronoi 
cell corresponding to the center $P$. It is obvious that for the considered limit 
the average length of Voronoi cells is $\langle d \rangle=1/n$.

In order to get the distribution function $g(d)$ of the Voronoi cell's length, first the 
distribution function $h(s)$ for the lengths between centers will be determined. Let us
start from the well-known Poisson distribution $P(N,t)$, giving the probability that inside
a length $t$ there are $N$ centers:  
\begin{equation}
P(N,t)=\frac{1}{N!}\langle N \rangle_t^N exp(\langle N \rangle_t).
\end{equation}
In the above equation $\langle N \rangle_t=nt$ stands for the expected (average) number of centers
on a length $t$. The probability that in an interval of length $t$ situated on the immediate right 
of $P$ there are no other centers is:
\begin{equation}
P(0,t)=exp(-nt).
\label{P0}
\end{equation}
The cumulative distribution $P_r(d_r>t)$ that the first neighbor at the right is at a distance 
$d_r$ bigger than $t$ is $P_r(d_r>t)=P(0,t)$.
The distribution function $g_r(d_r)$ for the lengths $d_r$ can be thus calculated as:
\begin{equation}    
g_r(d)=-\frac{\partial P_r(d_r>d)}{\partial d}=n \: exp(-nd).
\label{centers}
\end{equation}
Due to symmetry constraints, the same distribution function should apply for the $d_l$
lengths relative to the first neighbor in the left direction. The distribution function 
for the half of these intervals ($z=d_r/2$ or $z=d_l/2$) is given as:
\begin{equation}
w(z)=2n \:exp(-2nz).
\label{half}
\end{equation}
The length $d$ of the Voronoi cell is $d=\frac{d_l}{2}+\frac{d_r}{2}$,
and it's distribution $g(d)$ can be calculated as the convolution of two 
distributions of form (\ref{half}):
\begin{equation}
g(d)=\int_0^d w(z) w(d-z) dz=4n \:exp(-2nd). 
\label{length}
\end{equation}

\begin{figure}
\begin{center}
\epsfysize=80mm
\epsffile{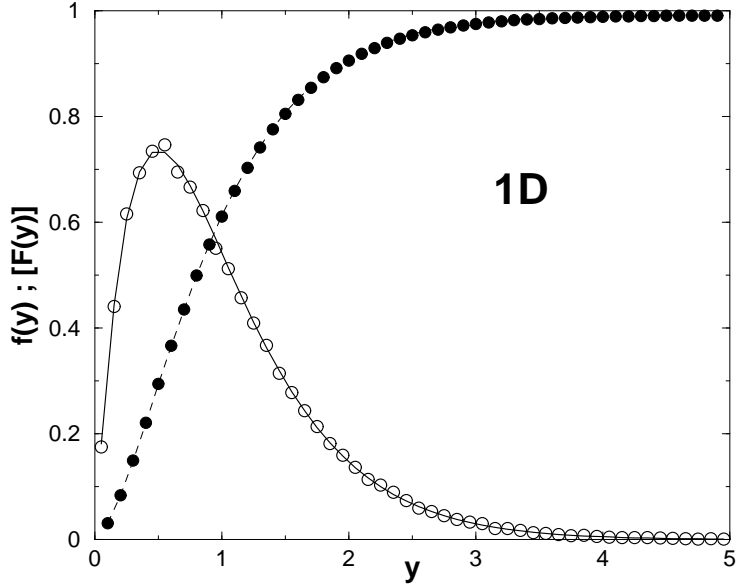}
\caption{Simulation results (empty circles) in 1D in comparison with the (\ref{1d}) exact result
(solid line). Results for the cumulative 
distribution function are also plotted. Filled circles are simulation data and the dashed line
is given by equation (\ref{F1}).}
\label{fig4}
\end{center}
\end{figure}

It is immediate to realize that this distribution function is normalized for 
$L \rightarrow \infty$. The distribution function for the adimensional 
quantity $y=d/\langle d \rangle$ is given then as 
\begin{equation}
f_{1D}(y)=4y \: exp(-2y),
\label{1d}
\end{equation}
which has the general form (\ref{fit}) with $a=b=2$. 
The cumulative distribution function $F_{1D}(y)$ is given by
\begin{equation}
F_{1D}(y)=1-(2y+1)e^{-2y},
\label{F1}
\end{equation}
and the moments of $f_{1D}(y)$ are immediately calculable:
$\langle y \rangle_{1D} =1;\;\langle y^2 \rangle_{1D} =\frac{3}{2};\;\langle y^3  \rangle_{1D} = 3.$
The most probable normalized length obtained from (\ref{1d}) is $y_{1D}=\frac{1}{2}$.

A simple computer simulation exercise can easily convince us about the validity of our calculations.
Results in this sense are presented on Figure \ref{fig4}. As an interesting observation one can realize that the distribution function for the lengths between randomly displaced centers (given by (\ref{centers}) ) is qualitatively different from the (\ref{length}) distribution function for the length of 
Voronoi cells (see also \cite{fortes}).   

\begin{figure}
\begin{center}
\epsfysize=100mm
\epsffile{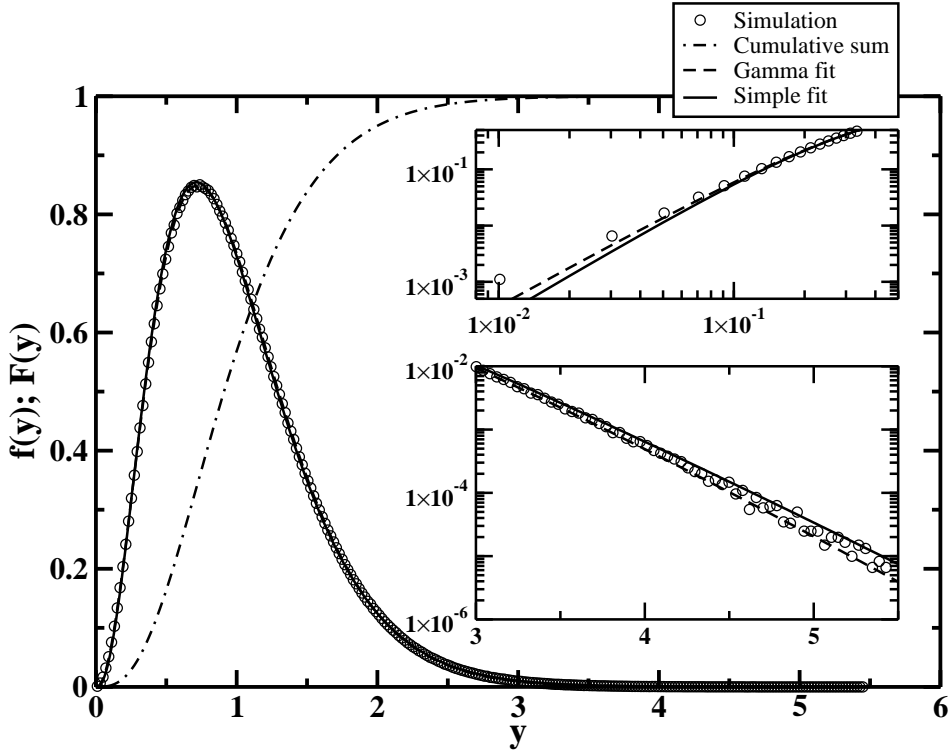}
\caption{The Voronoi cell's normalized area-distribution function in 2D. 
Empty circles are simulation results and the solid line is the (\ref{2d}) formula suggested 
in this study. 
With point-dashed line the cumulative distribution function, and with dashed line 
the (\ref{gamma}) best gamma 
function fit is plotted. The two inset figures are magnification of the 
small and large $y$ limits plotted on log-log 
and log-normal scales, respectively. 
On the scale 
of the figure there is no detectable difference between the cumulative 
distribution function 
calculated from simulation and the analytical expression given by (\ref{gamma}) and (\ref{2d}).}
\label{fig5}
\end{center}
\end{figure}

\section{The two-dimensional case}

Theoretical attempts to get analytical result for $f_{2D}(y)$ ($y=S/\langle S \rangle$, 
with S the area of Voronoi cells) in 2D, failed. 
We considered thus Monte Carlo-type computer 
simulations and fitted our simulation data in different forms. 
In particular, we focused on a 
three parameter fit in the generally accepted (\ref{gamma}) form and tried also 
a simple two parameter approximation (\ref{fit}) with handy $a=b$ values.  
It was found that the simple choice $a=b=7/2$ gives a visually good fit. For the 
normalized distribution function of Voronoi cell areas in 2D we proposed thus the
\begin{equation}  
f_{2D}(y)=\frac{343}{15} \sqrt{\frac{7}{2 \pi}}  y^{\frac{5}{2}} exp(-\frac{7}{2} y) 
\label{2d}
\end{equation}
simple approximation. On Figure \ref{fig5} we plotted with a continuous line the curve (\ref{2d}) in comparison 
with simulation data obtained on $29.889 \times 10^6$ Voronoi cells (almost three times more than the number of cells used by Tanemura). On the same graph it is drawn with dashed line the best gamma function (\ref{gamma}) fit. 
The $F_{2D}(y)$ cumulative distribution function is also plotted with dash-dotted line.  

At a first glance there is no detectable difference between computer simulation results, the curve suggested by
(\ref{2d}) and the gamma-function fit. Magnifying however the initial part and tail 
of the distribution function and plotting it on log-log 
and log-normal scales (insets in fig. \ref{fig5}), respectively,
one can observe slight differences. As expected, the three parameter gamma-function fit 
is better, but the improvement relative to (\ref{2d}) is not spectacular. The best-fit parameters obtained by us
for (\ref{gamma}) are $a=2.2975, b=3.01116$ and $c=1.0825$, in comparison with the values
$a=3.315$, $b=3.04011$ and $c=1.078$ obtained by Tanemura \cite{Tanemura2004,Tanemura2003}. 
For the analytically known second moment of the distribution ($\langle y_{theor2D}^2 \rangle=1.280$) 
our simulation data gives $\langle y_{sim2D}^2 \rangle=1.28231$, and the three-parameter gamma-fit yields $\langle y_{gamma2D}^2 \rangle=1.27947$. 
The error relative to the exactly known result is of the same order ($0.04\%$) as in the case of the fit given by Tanemura to his own computer simulation results. 
 
Using (\ref{2d}) all the important moments can be analytically calculated:
$\langle y \rangle_{2D} = 1;\;\langle y^2 \rangle_{2D} = \frac{9}{7};\;\langle y^3 \rangle_{2D} = \frac{99}{49}$.
The second moment has of course a much bigger relative error ($0.4\%$) respective to the exactly known value than the one obtained with the more sophisticated three-parameter gamma-function fit. 
This relative error is however still quite small and usual experimental data on Poisson Voronoi type patterns 
gives deviations of the order of a few percents. The most probable normalized area is $y_{prob2D}=\frac{5}{7}$.

\section{The three-dimensional case}

Due to the complex geometry involved, the possibility to analytically calculate $f_{3D}(y)$ 
($y=V/ \langle V \rangle$, V the volume of Voronoi cells) in 3D is even more gloomy. 
We performed thus again large scale computer simulations, studying the statistics of
$18.27 \times 10^6$ Voronoi cells (six times more than the statistics considered by Tanemura).
The three-parameter gamma-function gives a good fit for the simulation data, but again as in the
two-dimensional case a simple fit of the form (\ref{fit}) works also reasonably well and handy
$a=b=5$ values can be considered. We suggest thus that in 3D the Voronoi cell's 
normalized volume distribution can be approximated as:

\begin{equation}
f_{3D}(y)= \frac{3125}{24} y^{4} exp(-5 y) 
\label{3d}
\end{equation}

\begin{figure}
\begin{center}
\epsfysize=100mm
\epsffile{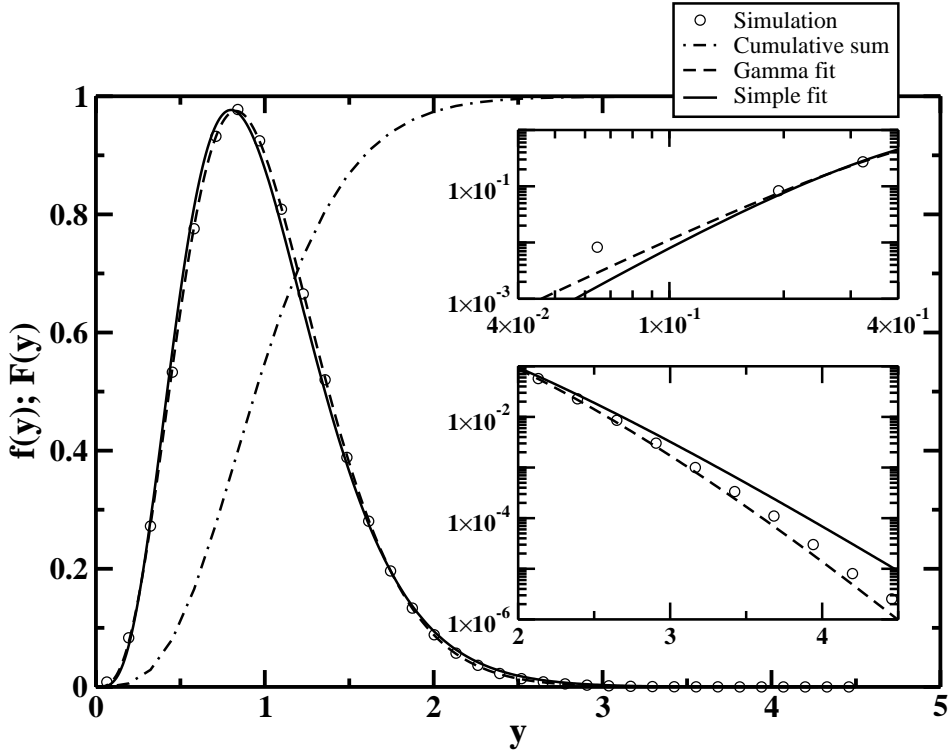}
\caption{The Voronoi cell's normalized volume-distribution function in 3D. 
Empty circles are simulation results and the solid line is the (\ref{3d}) formula suggested 
in this study. 
With point-dashed line the cumulative distribution function, and with dashed line 
the (\ref{gamma}) best gamma 
function fit is plotted. The two inset figures are magnification of the 
small and large $y$ limits plotted on log-log 
and log-normal scales, respectively. 
On the scale 
of the figure there is no detectable difference between the cumulative 
distribution function 
calculated from simulation and the analytical expression given by (\ref{gamma}) and (\ref{3d}).
}
\label{fig6}
\end{center}
\end{figure}

On Figure \ref{fig6} simulation results (empty circles) are compared with the (\ref{3d}) approximation (continuous line) and the three parameter gamma-function fit (dashed-line). In a first visual approximation one will realize that both curves describe well the simulation data. Magnifying however the initial part and tail 
of the distribution function and plotting it on log-log and log-normal scales (insets in Figure \ref{fig6}), respectively, one can observe the differences. As expected, the (\ref{gamma}) gamma-function
fit is better, and follows more the trend of the simulation results. The best fit parameters obtained in this study are $a=3.24174$, $b=3.24269$ and $c=1.26861$ (in contrast with $a=4.8065$, $b=4.06342$ and $c=1.16391$ found by Tanemura \cite{Tanemura2004,Tanemura2003}). The improvement relative to the simple (\ref{3d}) approximation is however again not spectacular, and is relevant only in the limit of very large or very small Voronoi cells. These limits does not appear usually in real experimental data, due to the fact that a much weaker statistics is achieved (patterns with less than $10^4$ cells are studied). On the figure it is also plotted (point-dashed line) the form of the cumulative distribution function $F_{3D}(y)$. On the scale of the image there is no detectable difference in the
cumulative distribution function determined from simulation and the forms (\ref{3d}) or (\ref{gamma}).

By using (\ref{3d}) the important moments are analytically calculable:
$\langle y \rangle_{3D} = 1;\;\langle y^2 \rangle_{3D} = \frac{6}{5};\;\langle y^3 \rangle_{3D} = \frac{42}{25}$. 
The most probable normalized volume is $y_{prob3D}=\frac{4}{5}$. For the second moment the relative 
error respective to the analytical exact results ($\langle y_{theor3D}^2=1.18$) is $\epsilon=1.7\%$. 
The gamma-fit for the simulation data yields $\langle y_{3Dgamma}^2 \rangle = 1.18683$ ($\epsilon=0.57\%$) while Tanemura's fit seems better yielding $1.17830$ ($\epsilon=0.14\%$). The second moment computed directly from simulation data is $\langle y_{sim3D}^2 \rangle=1.19$, giving the $\epsilon=0.85\%$ relative error.  

In agreement with the simulations of Tanemura \cite{Tanemura2004,Tanemura2003} we have also found that the
values $a=b=6$ suggested by Kiang \cite{Kiang1966} are not appropriate and give no good fit to our simulation data.

\section{Conclusions}
  
Motivated by the simple form of the exact result (\ref{1d}) for the size distribution function of Poisson Voronoi cells in 1D we proposed simple expressions for approximating the distribution in 2D and 3D where no exact results are available.  Exceeding the statistics considered in all previous studies computer simulations were used to investigate numerically the distribution function. It was shown that a simple 
form (\ref{fit}) with $a=b$ is appropriate for all practical applications to approximate the size-distribution of the
Poisson Voronoi cells. In 1D the exact results gives $a=2$. In 2D and 3D we found that $a=7/2$ and $a=5$, respectively, gives fair approximation. The simple values suggested for $a=b$ allows also to write in a compact form the approximations (\ref{2d}) and (\ref{3d}). If we denote by $d$ the dimensionality of the problem ($d=1,2,3$), the value of $a$ can be given as $a=(3d+1)/2$. Equations (\ref{1d}),(\ref{2d}) and (\ref{3d}) can be written then in a compact form as:
\begin{equation}
f_d(y)=\frac{((3d+1)/2)^{(3d+1)/2}}{\Gamma((3d+1)/2)} y^{\frac{3d-1}{2}} exp(-\frac{3d+1}{2}y)
\label{compact}      
\end{equation}
This distribution function is not an exact one and it is less accurate than a more complicated three parameter fit given
by the generalized gamma-function. Mathematicians will probably not appreciate it... but due to it's simplicity it will definitely be of importance for experimental scientist studying and characterizing complex Voronoi diagram-like patterns. 

{\underline{Acknowledgment}}
This work was supported by the Romanian CNCSIS 41/183 research grant. The research of F. J\'arai-Szab\'o has also been supported by the Felowship Program for Transborder Hungarian Scientific Research - Hungarian Academy of Science.


\begin{thebibliography}{10}

\bibitem{book} A. Okabe, B. Boots, K. Sugihara and S. N.  Chiu, {\it Spatial Tessellations: Concepts and Applications of Voronoi Diagrams} (Wiley, Chichester, 2000)
\bibitem{deberg} Chapter 7 (Voronoi diagrams) in "Computational Geometry: Algorithms and Applications, Second Edition" by Mark de Berg, Marc van Kreveld, Mark Overmars, Utrecht (the Netherlands) and Otfried Schwarzkopf, Hong Kong (China) 
\bibitem{meijering1953} J.L. Meijering, {\it Philips Research Reports}, {\bf 8}, 270 (1953)
\bibitem{Drouffe1984} J. M. Drouffe and C. Itzykson, {\it Nucl. Phys. B} {\bf 235}, 45 (1984)
\bibitem{Jeraud} G.R. Jerauld, J. C. Hatfield and H. T. Davis, {\it J. Phys. C} {\bf 17}(9), 1519 (1984); 
G.R. Jerauld, L. E. Scriven and H. T. Davis, {\it J. Phys. C} {\bf 17}(19), 3429 (1984)
\bibitem{DiCenzo1989} S. B. DiCenzo and G.K. Wertheim, {\it Phys. Rev. B} {\bf 39}(10), 6792 (1989)
\bibitem{Winterfield1981} P. H. Winterfield, L. E. Scriven and M. T. Davis, {\it J. Phys. C} {\bf 14}, 2361 (1981)
\bibitem{Talmon} Y. Talmon and S. Prager, {\it Nature} {\bf 267}, 333 (1977); 
Y. Talmon and S. Prager, {\it J. Chem. Phys.} {\bf 69}(7), 2984 (1978)
\bibitem{Brumberger1983} H. Brumberger and J. Goodisman, {\it J. App. Cryst.} {\bf 16}, 83 (1983)
\bibitem{Yoshioka1989} S. Yoshioka and S. Ikeuchi, {\it Ap. J.} {\bf 341}, 16 (1989)
\bibitem{Crain1976} I. K. Crain, {\it Random Processes in Geology} (New York: Springer-Verlag, 1976)
\bibitem{bio} A. Okabe and A. Suzuki, {\it Eur. J. Oper. Res.} {\bf 98}, 445 (1997); 
S. C. M\"uller, T. Mair and O. Steinbock, {\it Biophys. Chem.} {\bf 72}, 37 (1998); 
M. Tanemura, H. Honda and A. Yoshida, {\it J. Theor. Biol.} {\bf 153}, 287 (1991)
\bibitem{ecology} W. D. Hamilton, {\it J. Theor. Biol.} {\bf 31}, 295 (1971); 
M. Tanemura and M. Hasegawa, {\it J. Theor. Biol.} {\bf 82}, 477 (1980)
\bibitem{Boots} B. N. Boots, {\it Geografiska Annaler} {\bf 55B}, 34 (1973); 
B. N. Boots, {\it The Canadian Geographer} {\bf 19}, 107 (1975)
\bibitem{rivier} N. Rivier, in Disorder and Granular Media, edited by D. Bideau and A. Hansen (Elsevier Science Publishers B.V., New York, 1993)
\bibitem{lecaer} G. Le Caer and R. Delannay, {\it J. Phys. I France} {\bf 3}, 1777 (1993)
\bibitem{weaire} D. Weaire and N. RiviFortes M A and Pina P 1993 Phil. Mag. B 67 263 .er, {\it Contemp. Phys.} {\bf 25}, 59 (1984) 
\bibitem{Tanemura2003} M. Tanemura, {\it Forma} {\bf 18}, 221 (2003)
\bibitem{Tanemura2004} M. Tanemura in {Proceedings of Intersections of Art and Science, July 8-14, 2001, Sydney} eds. G. Lugosi and D. Nagy (http://www.mi.sanu.ac.yu/vismath/proceedings/)
\bibitem{Hinde1980} A. L. Hinde and R. E. Miles, {\it J. Stat. Comp. Simul.} {\bf 10}, 205 (1980)
\bibitem{kumar} S. Kumar, S.K. Kurtz, J.R. Banavar and M.G. Sharma, {\it J. Stat. Phys.} {\bf 67}, 523 (1992)
\bibitem{Weaire1986} D. Weaire, J. P. Kermode and J. Wejchert, {\it Phil. Mag. B} {\bf 53}(5), L101 (1986)
\bibitem{Kiang1966} T. Kiang, {\it Zeitschrift f\"ur Astrophysik} {\bf 64}, 433 (1966)
\bibitem{gilbert} E.N. Gilbert, {\it Annals of Math. Statist.} {\bf 33}, 958 (1962)
\bibitem{fortes} M.A. Fortes and P. Pina, {\it Phil. Mag.} B {\bf 67}, 263 (1993).
		     
\end{thebibliography}
\end{document}